# MetaStates: An Approach for Representing Human Workers' Psychophysiological States in the Industrial Metaverse

**AITOR TOICHOA EYAM[1], JOSE L. MARTÍNEZ LASTRA[1], Member, IEEE**

[1] FAST-Laboratory, Faculty of Engineering and Natural Sciences, Tampere University, 3320 Tampere, Finland

Corresponding author: Aitor Toichoa Eyam (e-mail: aitor.toichoaeyam@tuni.fi).

This research was supported by AI Powered human-centred Robot Interactions for Smart Manufacturing (AI-PRISM) project. The AI-PRISM project has received funding from the European Union's Horizon Europe research and innovation program under grant agreement No. 101058589.

**ABSTRACT** Photo-realistic avatar is a modern term referring to the digital asset that represents a human in computer graphic advanced systems such as video games and simulation tools. These avatars utilize the advances in graphic technologies in both software and hardware aspects. While photo-realistic avatars are increasingly used in industrial simulations, representing human factors such as human workers' psychophysiological states, remains a challenge. This article addresses this issue by introducing the concept of MetaStates which are the digitization and representation of the psychophysiological states of a human worker in the digital world. The MetaStates influence the physical representation and performance of a digital human worker while performing a task. To demonstrate this concept, this study presents the development of a photo-realistic avatar enhanced with multi-level graphical representations of psychophysiological states relevant to Industry 5.0. This approach represents a major step forward in the use of digital humans for industrial simulations, allowing companies to better leverage the benefits of the Industrial Metaverse in their daily operations and simulations while keeping human workers at the center of the system.

## I. INTRODUCTION

Industry 4.0 combines technologies such as Industrial Cyber-Physical Systems (CPS), and the Industrial Internet of Things (IIoT) to improve the efficiency and productivity of the manufacturing industry by combining the physical and digital worlds [1]. Employing this context as a foundation, Industry 5.0 complements Industry 4.0 by enhancing the focus on the development of more human-centric and sustainable solutions [2].

Among a broad spectrum of technologies that have experienced a huge advance during the last twenty years, simulations play a crucial role in the development of applications in Industry 5.0. Simulations allow the development of digital twins, which are digital replicas of physical systems. By utilizing digital twins, it is possible to apply real-time monitoring, run better analysis, and optimize processes, being able to enhance the sustainability of the processes [3]. At the same time, simulations enable the possibility of tailoring processes to human needs and provide safe spaces for workers to upskill and reskill their abilities [4].

The physical-digital dyad that enables the integration of both realms finds its conception in the Metaverse. Introduced by *Neal Stephenson* in his book *Snow Crash,* the Metaverse represents a virtual interconnected space where physical and digital entities merge interacting with each other and exchanging information [5]. Even though the creation of the Metaverse is still in its early stages, companies such as *Nokia,* and *Microsoft,* have already seen three potential subdivisions [6], [7]. A Consumer Metaverse (CM) focused on socially immersive experiences based on entertainment and leisure. An Enterprise (or Commercial) Metaverse (EC) that intends to apply immersive technologies into the business environment to design products, and improve communications and collaboration. Finally, the Industrial Metaverse (IM) aligns its goals with Industry 5.0, aiming to blend physical and digital realities making humans and artificial intelligence (AI) collaborate to improve productivity, safety, and sustainability.

While physical systems such as machines or robots are being replicated into IM applications, it is still a challenge to integrate human workers in three-dimensional simulations. Although several industrial applications contemplate digital human in their simulations, they are mainly considering just their body structure and range of motion. Even though such characteristics have been widely studied to assess the ergonomics of work operations, these simulations rarely consider key aspects that shape human behavior and performance such as their psychophysiological states [8]. Moreover, current virtual training environments also fall short of integrating and representing the psychophysiological states that workers are experiencing during their training [9]. Observing and understanding these human states becomes even more relevant in collaborative scenarios where workers have to interact with each other.

Considering that psychophysiological states have a direct impact on human behavior, communication, learning, and performance [10], it is crucial to achieve proper ways of expressing and representing these states in a simulation environment. Thus, this paper aims to address the existing gap in the current state of the art by exposing an approach for representing psychophysiological states in industrial simulations. This approach can be applied in immersive and non-immersive virtual scenarios, enabling the understanding of human workers' states to subjects immersed in virtual realities, and professionals running desktop-based application environments. To accomplish this objective, the following research question is considered in this paper:

- RQ1: How to represent the psychophysiological states of a human worker in a digital human?

Overall, the remaining part of the paper proceeds as follows. Section II describes the literature review concerning this study. Section III presents the main concepts employed in this work to address the gap in the current literature. Section IV illustrates how the concepts described in the previous section can be implemented in an industrial use case. And lastly, Section V summarizes the goals achieved in this document and explores future work in the research field.

## II. PREVIOUS WORK AND STATE OF THE ART

Introducing realistic representations of human workers in the Metaverse requires research on the concepts and methods in the domain. Such groundwork is presented as follows: subsection A focuses on the Metaverse and its future in industrial applications, subsection B shows the current applications of avatars in the industry, and subsection C presents the current state of representing human states in digital applications.

### A. INDUSTRIAL METAVERSE: THE FUTURE OF INDUSTRIAL APPLICATIONS

The IM rises as a subsection of the Metaverse which focuses on industrial applications such as manufacturing, construction, or logistics [7], [11]. The IM aims to make humans and AI entities work together to design, build, operate, and optimize physical systems using digital technologies [11], [12]. To achieve these goals different technologies are combined, including the Internet of Things (IoT), big data, digital twins, extended reality (XR), and AI models [13], [14].

**TABLE 1.** Benefits of the Industrial Metaverse for Industry.

| Title | Description | Benefits |
|---|---|---|
| Enhanced Human-AI Interactions | Humans and AI can work together in designing, building, operating, and optimizing physical systems | Improve efficiency and innovation. Humans and AI mutual learning |
| Technology Integration | Easy integration of technologies such as IoT, big data, digital twins, XR, and AI Models in the same ecosystem | Better decision-making and resource utilization. Better data generation |
| Simulation-Driven Optimization | Enables data integration from multiple sources, and adjustment of simulation variables | Better process definition, production efficiency, design, and sustainability |
| Remote Monitoring and Real-Time Visualization | Facilitation of visualizing real-time data processes and remote monitoring capabilities | Faster identification of problems and decision-making |
| Training and Maintenance in Simulated Environments | Support of realistic platforms allowing XR settings for training and maintenance. Platforms supporting realistic virtual settings for training | Enhancement of workers' skills and ensuring safer operations. Better guidance and process visualization |
| Supply Chain Optimization | Leveraging digital twins, real-time tracking, and predictive analytics | More efficient logistics, reduced costs, and improved performance |

The development of an IM that facilitates human-AI interactions allows the AI models to learn from the human, and the human to learn from the AI models [7]. As far as simulations are concerned, they play a major role in the IM by enabling the integration of data from multiple sources with a common goal. In the IM, it is possible to improve aspects such as production efficiency or sustainability footprints by adjusting simulation variables of the industrial process [11]. Apart from production simulations, other key features of the IM include remote monitoring, real-time data visualization, or training and maintenance in simulated environments [15], [16]. The IM is evolving and being applied to several fields such as smart factories or supply chain. As regards smart factories, XR interfaces are used to assist human workers in visualizing production processes, provide them guidance, or train them in realistic virtual environments [16]. In the case of supply chain, the IM enhances its optimization through digital twins, real-time tracking, and predictive analytics [15]. TABLE 1 resents a summary of the mentioned benefits that the IM brings to the industrial field.

One of the emerging platforms that allows the integration of technologies to develop IM applications is the *NVIDIA Omniverse* [17]. Employed by several companies such as *BMW* or *Siemens* [18], [19], *NVIDIA Omniverse* facilitates the development of 3D workflows and applications. Overall, companies such as *Amazon*, *Mercedes-Benz*, *Nokia*, *Boeing*, and *Volkswagen Group* support the relevance of employing IM solutions in their systems for improving their results in the fields of factory simulation, mobile networks, training, construction planning, and project management [6], [20], [21], [22].

### B. AVATARS: DIGITAL HUMANS IN INDUSTRIAL SIMULATION ENVIRONMENTS

One of the most critical aspects of simulated environments is providing optimal immersion. A particular aspect to achieve the feeling of immersion is the realism that the simulation provides [23]. Consequently, to improve the subjective experience of humans in simulated environments, it is relevant to have realistic digital humans, or avatars, in them [24]. An avatar is the representation of the self in a digital world [5], [25]. As avatars, the users can interact with digital entities and explore the digital world in which they are immersed. Depending on the characteristics and target of the digital world, the appearance and features of the avatars may vary. In the gaming industry, avatars may differ completely from the self, even with the possibility of having a non-human avatar [26]. On the one hand, in social platforms, users might select avatars that reflect their personal interests, creativity, or desired persona, which can extend beyond human representation [27]. On the other hand, when talking about IM, it becomes relevant to have a more realistic representation of oneself in the digital world to enhance the user-avatar relationship [28]. The rise of the gaming industry, social platforms, and other simulation environments has incentivized the development of platforms to generate photo-realistic characters. Among these applications are *Autodesk Maya* [29], *Blender* [30], *Unreal Engine's (UE) MetaHuman* tool [31], [32], *Reallusion's Character Creator* [33], and *DazStudio* [34]. The use of photo-realistic avatars in digital worlds enhances the user experience by evoking deeper immersion levels and facilitating the connection with the application [35]. With regards to the appearance of avatars, companies such as *Meta*, and *Apple* are investing resources in generating photo-realistic digital humans of users to provide a deeper sense of realism and immersion in Metaverse applications [36], [37]. Even though the advances in digital human appearance have significantly improved over the years, eluding falling into the Uncanny Valley [38], little progress has been achieved in representing human internal and physical states in the IM. Examining industrial simulations, they commonly focused on the characteristics of the machinery, robots, or production allocation factors [39]. These elements are represented realistically, respecting their dimensions, colors, and other visual features. Also, industrial simulations enable the user to view and update internal features of machinery elements, such as speed, range of motions, and production time. Additionally, these types of applications make use of the physical aspect of digital humans focusing on their posture and motion. This is accentuated in the manufacturing sector where there can be found commercial solutions and research work in this area [8]. One of the most employed human simulation systems in engineering is *Jack,* which focuses on posture, accessibility, and reachability assessments [40]. Another commercial solution employed in the manufacturing sector is *UMTRI Human Shapes*. This software allows users to intuitively generate different body shapes applied to automotive interior and crash assessment [41]. As indicated in the reviews of [8] and [42], the research work using digital humans in this area mainly addresses ergonomics aspects. However, there is still work to be done such as considering additional human factors and developing more detailed digital humans that provide useful information to the simulations.

### C. HUMAN STATES IN DIGITAL APPLICATIONS

As introduced in the previous section, human beings are mainly represented in digital applications by their physical capacities [8]. Depending on the application, digital humans could have other qualities that allow them to interact and exchange information with the digital environment in different manners. As an example, in the gaming industry, avatars have a set of attributes that represent their statistics, states, or capabilities. For example, in role-playing games (RPGs), a character could have *physical* attributes such as *strength, stamina,* or *speed*, and *mental* attributes such as *intelligence, wisdom,* or *charisma* [43], [44]. These attributes enable the option of generating avatars with similar appearance but with different capacities. A great example of how attributes or states can influence how an avatar interacts with the surrounding digital world is depicted in *The Sims* videogame [45], [46]. *The Sims* presents a "real-life" simulation in which digital humans have multiple physical and mental needs that influence their actions and behavior. Depending on the level of those physical and mental states that the character is experiencing, they will build skills faster or slower, or their interactions with other characters might be more successful. In video games such as *The Sims 4* the representation of physical and mental states has been performed employing color patterns and displaying information in the user interface (UI). For example, emotional states such as *happy*, *sad, angry,* or *tense,* were represented by displaying the colors *green, dark blue, red,* and *peach*, respectively, in the UI [47], [48].

Another example of representing human physical and mental states in digital environments was presented in the work undertaken in [49]. The authors in [49] proposed an approach for representing emotional states in avatars by growing or contracting their fur or by adjusting the brightness of a particle system. Similarly, the authors in [50] represented physiological signals related to emotional and cognitive states in avatars with *skeuomorphic, particles, creatures,* and *environmental* visualizations. As seen in the examples, the gaming industry has made great efforts to represent human states in their platforms. This objective stems from the desire to evoke emotions and cognitive states in their users generating remarkable experiences. On the other hand, industrial applications have traditionally had objectives such as productivity, efficiency, and safety. Thus, the more prominent human factors considered in industrial applications are related to ergonomics (e.g.: body structure, height, weight, or age). There are very few studies that combine digital humans and human states in industrial scenarios. The work performed by [51] proposes a simulation for time allowances that integrate workers' fatigue, learning, and reliability as main human factors. In addition, even though the research conducted in [52] is also focused on ergonomics, they employ a standard to analyze human workers' fatigue over a cycle of the operation. Although both examples take into consideration a psychophysiological state such as fatigue to influence the results of the simulation, their research does not include visual representations of the states. The lack of a visual representation in an IM with multiple entities interacting with each other makes it more difficult to understand the status of human workers in the simulation. Areas such as virtual training, real-time monitoring, and simulation algorithms can benefit from including visual feedback on human states. Therefore, there is a need for further research on the integration and representation of psychophysiological states in industrial simulations.

## III. METAHUMANS & METASTATES: HUMAN WORKERS INSIDE THE INDUSTRIAL METAVERSE

This research aims to represent the psychophysiological state of human workers in the IM. To achieve an optimal representation of a human worker in the digital world, it is crucial to define which main aspects are needed to represent a human worker in the workplace. This research work highlights as the main characteristics of a digital representation of a human worker: physical appearance, and psychophysiological states. The following subsections discuss how to meaningfully represent human workers' physical appearance and psychophysiological states in the context of the IM.

### A. METAHUMAN

The goal of this study is to enhance the representation of a human worker in simulations by including their psychophysiological states. Consequently, this research presents the concept of MetaHuman as the digital representation of a human worker in the IM. Therefore, a MetaHuman includes a photo-realistic representation of the person it portrays. Photorealism refers to their main physical features (facial, skin, body proportions), as it is crucial that visually the operators can be recognized as themselves. Furthermore, a MetaHuman goes beyond the visual representation, including internal traits of the person it represents, such as their emotions and cognitive states. As the context of this approach is inside the IM, the psychophysiological characteristics of a MetaHuman mainly target the response of the human worker while performing specific tasks. By combining the external and internal aspects of a human worker, a MetaHuman transforms the role of a digital operator from an illustrative entity to another variable to be considered when running simulations of industrial processes. As a result, three main goals are achieved, increasing the user experience, enhancing the realism of the simulation, and enabling the possibility of having better algorithms that introduce psychophysiological factors.

### B. METASTATES

The psychophysiological states are internal processes that may influence how human workers perform operations and tasks [53]. To perceive how a human worker might be feeling while performing a task in a simulation, it is important to generate a method of illustrating those internal states. Similar to the concept of MetaHuman referring to the digital representation of a human worker, this study introduces the concept of MetaStates. The word MetaStates comes from the union of two words, *meta* and *states*. *Meta*, a Greek prefix meaning "beyond", is used to indicate an abstraction of a concept and to complete it. *States*, on the other hand, refer to the multiple ways of being of a person, which– can result in physical, emotional, and cognitive conditions. Thus, this work defines the MetaStates as the representation of the psychophysiological states of a human worker in the digital world.

Taking into consideration the context of manufacturing in Industry 5.0, some of the most relevant human factors to consider are *stress, attention, cognitive workload,* and *physical fatigue* [53]. These states find their depiction and computation in a simulation environment as MetaStates. A human has psychophysiological states and a MetaHuman has MetaStates. The MetaStates quantify and represent the psychophysiological states that a human worker experiences while performing a task. This allows to understand the status of an operator while performing a task in a simulation environment.

The concept of MetaStates introduces a new layer of communication, allowing the expression of inner sensations, and enabling the possibility of making adaptations to the process, tasks, or performed operations to improve workers' state. A MetaState can be depicted in multiple levels of detail that take into consideration the characteristics of a MetaHuman and what their role in a simulation is. Inside a simulation environment, a MetaHuman performs the same activities that the analog human worker would perform in the real world. Therefore, the majority of the activities concerning MetaHumans are movements and actions. Consequently, MetaStates heavily influence MetaHuman's appearance, mobility, and performance. As it happens with humans, each person expresses differently an internal state. On one hand, some individuals experiencing physical fatigue might have an exhausted look or even require more time to make certain operations. On the other hand, there might be some individuals who might not express fatigue in their looks, but with their body language, and this may impact differently their task performance. The diverse variations of expressing internal states lead to the need to develop a MetaState Reaction Model (MRM) for each MetaHuman.

The MRM is based on the reactions of each human worker to the different psychophysiological states experienced during the task performance. When creating a MetaHuman, a basic MRM composed of common reactions towards an input is associated with it. Following, the more data is fed into the system, and the more information is known about the worker the MRM becomes more complete, obtaining a MetaHuman more accurate to the real person.

As mentioned earlier, the MetaStates aim to recreate the internal states of a human worker while performing a task, influencing on the mobility, and appearance of the MetaHuman. Considering that the aims of these applications are to enhance user experience, improve real-time feedback, and perform better simulations, there should be generated a visual clue for the workers running the simulation, or users experiencing the digital world in XR, which indicates the overall status of the MetaHuman. In this work, this status is presented as the MetaState Performance Index (MPI). The MPI represents the overall status of the MetaHuman to perform a given task, taking into consideration all the different MetaStates and the task to be performed. Inside the simulation environment, the MPI is represented by two main aspects, the MetaState aura, a circle generated in the base of the MetaHuman, and the MestaState sphere, a sphere located at the top of a MetaHuman's head. The idea behind using two elements to represent the MPI is to improve visual perception. This is because in a 3D simulation depending on the perspective, one of the elements can be hidden behind a third object (e.g., working table, robot, cell). With two ways of visualizing the MPI, the feedback to understand the status of the human worker is improved.

The MPI provides extensive information about the overall condition of the MetaStates in a simplified way. With respect to providing visual information to the worker running the simulation, the MPI updates its color as indicated in TABLE 2.

The values of the MetaStates can be achieved by capturing diverse psychophysiological signals of human workers via wearable devices.

TABLE 2. MetaState Performance Index color description.

| Color | Description |
| --- | --- |
| Green | The human worker is in an *optimal* status to perform the task up to their skill level. At this status, the worker is usually performing correctly or is exceeding performance. |
| Amber | The status of the human worker is a *threat* to performing the task up to their skill level. At this status, the person's overall performance decreases from optimal. |
| Red | The human worker is experiencing a *suboptimal* status to perform the task up to their skill level. In this situation the worker is not in good condition, thus it might need a break. |

As described in the previous work done in [53], the majority of the psychophysiological signals are controlled by the autonomic nervous system (ANS). The ANS regulates crucial body processes such as brain, cardiovascular, electrodermal, muscular, respiratory, and ocular activities. By employing wearable devices able to apply techniques such as electroencephalography (EEG), electrocardiography (ECG), or eye tracking, it is possible to extract features related to different human states. This study focuses on how to represent the values of the MetaStates in a useful way in industrial applications, assuming that the capture and analysis of the MetaStates have been previously carried out. Thus, this research work pays special attention to psychophysiological states that are more relevant in industrial scenarios such as *stress*, *attention*, *cognitive workload*, and *fatigue*, instead of considering other human states such as happiness, joy, or sadness. Nevertheless, in TABLE 3 there are some examples of how psychophysiological states such as *stress*, *attention*, *cognitive workload*, and *fatigue* can be captured using the abovementioned techniques.

TABLE 3. Detection of psychophysiological states using different techniques (adapted from [53]).

| Psycho-physiological State | EEG | ECG | Eye Tracking |
| --- | --- | --- | --- |
| Stress | Increased right hemisphere activity | Increased heart rate | Increased pupil diameter |
| Attention | Increased high-frequency activity | Higher heart rate variability at rest | Longer gaze fixation. More frequent saccades |
| Cognitive Workload | Decreased theta and/or alpha band power. Increased beta and/or gamma band power | Increased heart rate. Decreased inter-beat interval | Changes in pupil diameter. Increased blinking duration |
| Fatigue | Reduced peak alpha frequency near the motor cortex | Significant increase in heart rate | Longer fixations. Slower saccades. Increased blink frequency |

Different human workers would have different levels of each feature corresponding to a human state. As an example, for the same stressful stimulus, multiple human workers could have distinct heart rate responses. This means that in a high-stress situation, the heart rate response of an individual could be lower than the heart rate response of another person. Therefore, to be able to compare the responses and generate representation models, the MetaStates' values of a worker are classified into three ranges, *high*, *mid*, and *low*. These ranges are established after recording and studying the psychophysiological signals of a worker while they repeat the same task multiple times. Defining these ranges provides a normalization of the data, being able to compare and generate common responses for values classified under the same category. In addition, the values of the MetaStates are related to a specific task and are quantified. This implies that the values of a MetaState range from positive to negative valence. As can be seen in TABLE 4, *optimal* levels of a MetaState are considered as positive for task performance; threat levels are handled as a risk to task execution; and finally, *suboptimal* levels are treated as dangerous for the human worker and task performance. Following this explanation, the color used for visualizing the MPI is determined based on the range of the MetaStates' values. Depending on the characteristics of the task being performed, the criteria on how to weigh the influence of each MetaState for the MPI may be different. For example, in a physically demanding task, *physical fatigue* may have a greater weight on task performance than other MetaStates. Thus, the evaluation of the MetaStates' weight should be assessed individually for each task. Therefore, as a general view, if the overall state of the MetaStates is *optimal* for task performance, the MPI color will be green; in case the overall levels of the MetaStates are a *threat* to task execution, the MPI color will be amber; lastly, if the MetaStates' balance is considered *suboptimal* to perform the task, the MPC color will be red. The colors green, amber, and red, have been purposely selected as they are commonly associated with positive, cautious, and alarm status. As a consequence, a person working with the simulation can easily understand the general state of a MetaHuman by looking at the MPI color.

TABLE 4. **MetaStates range levels.**

| MetaState | Optimal | Threat | Suboptimal |
|---|---|---|---|
| Stress | *Low* | *Mid* | *High* |
| Attention | *High* | *Mid* | *Low* |
| Cognitive Workload | *Low* | *Mid* | *High* |
| Physical Fatigue | *Low* | *Mid* | *High* |

It is crucial to emphasize the fact that the analog values captured from the wearable devices are employed to generate the *high, mid,* and *low* ranges for each individual. Furthermore, the same analog values are saved to be employed by other variables and algorithms that use human factors data in the simulation environment. In addition, apart from providing information about the performance, the MPI serves as an indicator in Metaverse applications to differentiate between a MetaHuman and a digital avatar not based on a real person, or with non-playable characters (NPCs).

To finalize this section, FIGURE 1 illustrates how the concepts of MetaHuman and MetaStates enhance the current state of the art of digital humans in industrial applications. The schema presents that the concept of MetaHuman is composed of two main aspects, a photo-realistic digital human (grey), and their MetaStates (purple). The current state of the art is mainly focused on the creation of a photo-realistic digital human, whereas this study provides a new realm, the MetaStates. The MetaStates provide information about the psychophysiological states of a human worker. Visual feedback about these states is given with the MPI and also influences the animations of the digital human with the MRM.

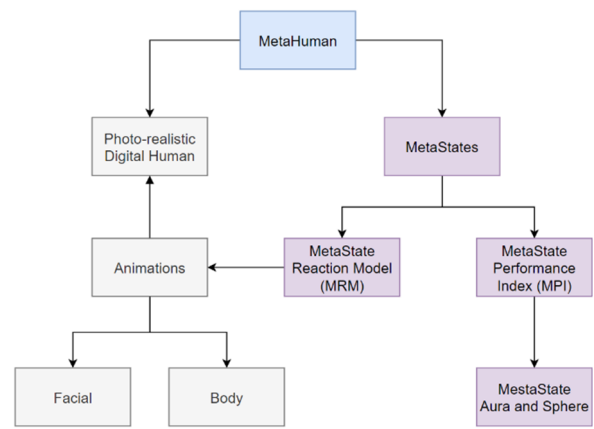

**FIGURE 1.** MetaStates influence in the state of the art.

## IV. IMPLEMENTATION

The previous section introduced the concept of MetaHuman as the digital depiction of a human worker, and its MetaStates, which are a representation of their psychophysiological states. This section exposes how the concepts of MetaHuman and MetaStates can be applied to a use case inside the manufacturing industry. To apply the concepts of MetaHuman and MetaStates inside the context of Industry 5.0, the implementation exposed in this section is a natural continuation of the work performed in [54]. The use case proposed in [54] presents the performance of a collaborative task between a human worker and a collaborative robot (cobot). Both actors, human and cobot, work collaboratively to fulfill the assembly of a wooden box which is comprised of six parts. During the performance of the assembly, the human worker wears an EEG headset which can detect different psychophysiological states. These psychophysiological states are sent to the cobot which updates parameters such as velocity and delay between operations adapting them to the worker's psychophysiological states. Therefore, this section proposes an approach for creating a photo-realistic digital human and employs the data generated in the mentioned use case to develop the visual representation of the MetaStates.

As exposed in the previous section, a MetaHuman is composed of two main aspects, a photo-realistic representation of a digital human, and the MetaStates of the digital human. FIGURE 1 presents a schema of the core elements that comprise the concept of MetaHuman in this research work. On the one hand, the photo-realistic digital human is nourished with facial and body animations to make the external representation of the human worker precise. On the other hand, the MetaStates are represented with the MPI and the MRM. Starting with the MPI, it takes into consideration all the MetaStates to determine if the worker is in an optimal condition to perform the task. To show that, the MPI updates the colors of the MetaState aura and sphere attached to the MetaHuman. Next, the MRM serves as a bridge between the MetaStates metrics and the animations that should be triggered to represent them.

Taking the schema presented in FIGURE 1 into consideration, the first input to develop a MetaHuman is generating a photo-realistic digital human. Several platforms and software can be used to generate the appearance of a MetaHuman. This research work utilizes *UE's MetaHuman Animator* and *Creator* for the generation of digital human workers. *UE's MetaHuman Animator* allows the generation of photo-realistic digital humans from a digital mesh of a human's face, or a face recording coming from a depth camera (head-mounted, or *iPhone 12* or newer). The approach selected for this research was using a depth camera from an *iPhone 12 Pro Max* to capture the facial expression of the human worker. Four main frames of his face were captured; frontal, right, and left frames with a relaxed neutral expression, and a frontal frame showing the teeth to register the bite (FIGURE 2).

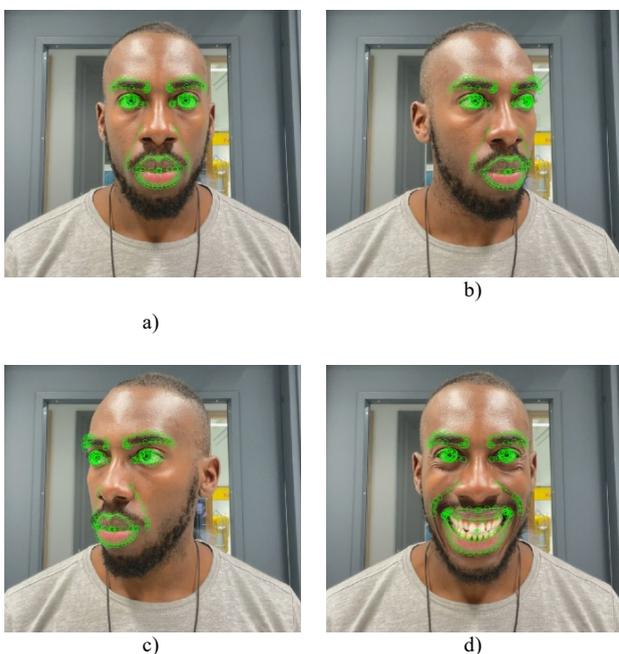

**FIGURE 2.** a) Frontal frame. b) Right frame. c) Left frame. d) Teeth frame.

Once the frames were extracted, and the features' trackers adjusted it was generated the first version of the digital human FIGURE 3.

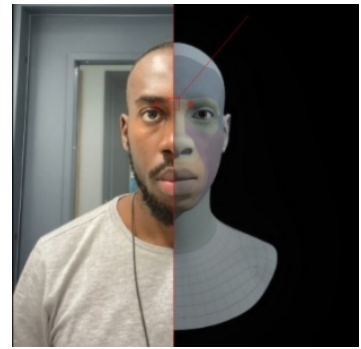

**FIGURE 3.** Human worker vs Digital worker mesh.

To achieve a better resemblance to the human worker, the digital mesh was adjusted using *UE's MetaHuman Creator*, updating characteristics such as face features, skin tone, eye color, hair, and body type. The result of the digital human can be seen in FIGURE 4, where it can be appreciated how current applications can be exploited to generate accurate photo-realistic digital humans for human workers.

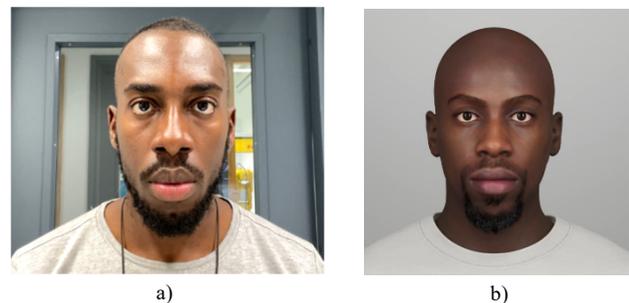

**FIGURE 4.** Human Worker vs Digital Human Worker: a) Human Worker, b) Digital Human Worker.

Once the photo-realistic digital human is generated following FIGURE 1, there should be also created animations that complement the external appearance of the human worker. For this input, it was also employed *UE's MetaHuman Animator*. As indicated in FIGURE 1, the animations performed by a MetaHuman are linked with the MetaStates that it is experiencing. As stated in the previous section, the MetaStates contemplated in this research work are of special interest to Industry 5.0. Accordingly, animations were created to represent two MetaStates, *stress* and *physical fatigue*. The reason behind selecting these MetaStates is because they are commonly expressed through facial and body movements.

Starting with *stress*, the generated animation was focused on the facial expression of the human worker. Similar to the creation of the digital human mesh, the human worker was recorded performing a stressed facial expression with an *iPhone 12 Pro Max*. Once the footage of the performance was defined, it was processed and mapped to the digital human mesh created in the previous step. The results of the processed animation can be seen in FIGURE 5. Regarding body animations, this research has integrated three main animations, idle, walking, and *physical fatigue*.

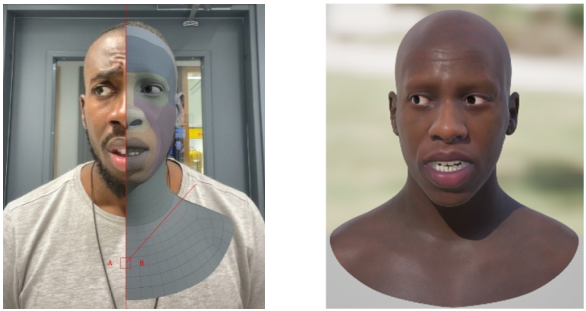

FIGURE 5. Stress performance vs Stress animation.

After finishing the creation of the photo-realistic digital human and its animations, the next step for creating a MetaHuman is integrating the representation of the MetaStates. To achieve this goal, it was developed a scene on UE staging a workstation composed of a MetaHuman, a cobot, and the wooden box to be assembled. To develop the MPI visualization, two dynamic elements were created, the MetaState aura, and sphere, which accompany the MetaHuman in all its actions.

In alignment with the previous section, both the aura and sphere adjust their color based on the MetaStates experienced by the MetaHuman. The threshold values of the MetaStates have been extracted from the psychophysiological data captured with the EEG headset in the use case experiment cited at the beginning of this section. The values of the metrics can range between 0 to 1, with 0 as the minimum and 1 as the maximum value. As the human worker performed the assembly of the box several times, it is possible to define the threshold limits for each MetaState by analyzing the metrics of the human worker. These allowed to determine which ranges of MetaState values could be mapped in the *optimal*, *threat*, or *suboptimal* interval. For example, the *optimal* range for the human worker's *attention* is *[0.51, 1]*, on the other hand, the *suboptimal* interval for *stress* is defined as *[0.64, 1]*. It is crucial to emphasize that the mentioned thresholds are valid for the human worker who has been conducting the experiments. In the case of other MetaHumans that are not based on this human worker, the numerical values would differ. Therefore, it is important to abstract the ranges into *high, mid,* and *low*, as specified in TABLE 4. This abstraction facilitates the comparison between MetaHumans and the development of representations and animations using the MPI and MRM.

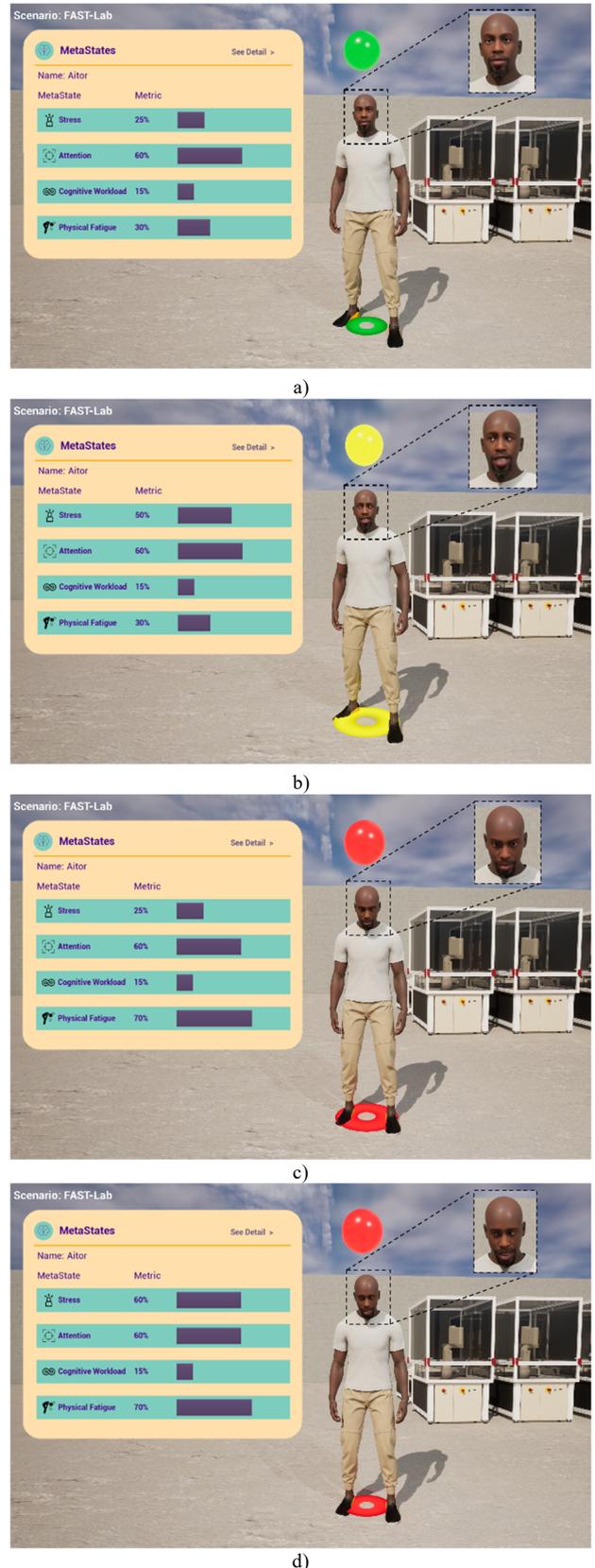

FIGURE 6. MRM animations and MPI visualization: a) MRM: idle – MPI: optimal, b) MRM: stress – MPI: threat, c) MRM: physical fatigue – MPI: suboptimal, d) MRM: stress and physical fatigue – MPI: suboptimal.

In the preceding section, it was explained that the weighting influence of the MetaStates on the MPI should be tailored to the task to be completed. In the scope of this implementation, the relevance of each MetaState has been assessed to be equally important regarding the task description. Therefore, it has been determined that when all the MetaStates are within their *optimal* range, the MPI color will be green. In case one or more MetaStates are a *threat* to performance, the color will turn amber. Lastly, if at least one MetaState falls into a *suboptimal* interval, the MPI color will change to red.

Finally, the MRM serves as a bridge between the MetaStates values, and the animations performed by the MetaHuman. In this scenario, two MetaStates were selected to trigger animations when they arrived at a certain value. The selected MetaStates are *stress*, which triggers a facial animation, and *physical fatigue* which triggers a body animation. On one hand, the facial animation is triggered when *stress* values are a threat to the task performance (FIGURE 6 b). In this case, the MetaHuman recreates the facial expression of the human worker when he experiences a stressful situation. On the other hand, the body posture animation is activated when *physical fatigue* is suboptimal for performing the task (FIGURE 6 c). In this animation the MetaHuman has a less upright posture than in the idle position, showing signs of fatigue. Following, FIGURE 6 d, presents a scenario in which multiple MetaStates are triggering reactions, in this case, the *stress* level is a thread and *physical fatigue* is suboptimal for fulfilling the task.

## V. CONCLUSIONS AND FURTHER RESEARCH

Creating simulations is a complex process that relies heavily on the accuracy of the introduced data. IM applications should take into consideration the premises of *Industry 5.0*, developing sustainable and human-centric solutions. To achieve this goal, it is crucial to properly introduce humans into the simulations considering how they can influence the process and how the process can affect them. This consideration will increase the realism of the simulations, their algorithms, and thus improving the outputs.

With the idea of generating human-centric IM applications, this study aims to provide a new approach to represent the psychophysiological states of human workers in the IM. To achieve this goal two concepts were presented, MetaHuman, and MetaStates. A MetaHuman was defined as the representation of a human worker in the IM, including a photo-realistic digital human and their MetaStates. Being the MetaStates the digital representation of the psychophysiological states of the human worker. The definition of these two concepts introduces a novelty in the field of industrial simulations expanding the state of the art, which was mainly focused on developing photo-realistic representations of persons, with little research done on how to include psychophysiological states in digital humans.

Even though the implementation of this study has been limited to exemplify how to represent MetaStates by employing the MPI, and animations triggered by the MRM, it serves as a foundation for further research on this topic. This research work opens a new era of IM applications which will be able to apply the concepts from this study to contemplate psychophysiological signals in their simulations. Consequently, future studies should focus on gathering human workers' data regarding their psychophysiological states to improve the ranges and algorithms of the MPI. Additionally, more research should be done to capture human workers' expressions and body language to generate better animations correlated with the MRM.

In conclusion, introducing MetaStates into IM applications can lead to more realistic simulations. These simulations can generate more sustainable and human-centric solutions by assessing how a situation can influence positively or negatively the well-being and performance of a human worker.

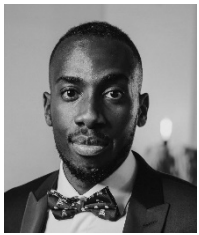

**AITOR TOICHOA EYAM** was born in Spain, in 1994. He received a B.Sc in Industrial Engineering from the Polytechnic University of Valencia, Spain in 2017. He made the M.Sc. specialization in MS Automation Engineering at Tampere University, Finland in 2019, and received a M.Sc. in Industrial Engineering with a specialization in Process Control, Automation and Robotics from the Polytechnic University of Valencia, Spain in 2020. Previously, he worked as an intern for the Engineering Department of Grupo Antolin Autotrim in Spain. During 2018-2019 he worked as a Research Assistant in the FAST-Lab research group, at Tampere University. Later in 2020, he worked as a software engineer at Capgemini's ADCenter in Valencia, Spain. Then, in 2021, he worked as a Researcher at the FAST-Lab research group at Tampere University. Currently, he is a Doctoral Researcher at Tampere University working on the project AI-PRISM from the Horizon Europe funding program. His main research interests include Human-AI Interactions, Robotics, Emotion-Driven technology, and Factory Automation.

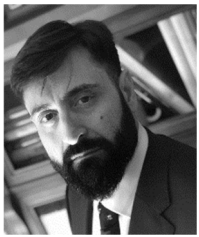

**JOSE L. MARTÍNEZ LASTRA** (Member, IEEE) received the Ingeniero Técnico Industrial degree in electrical engineering from the Universidad de Cantabria, Santander, Spain, and the M.Sc. degree (Hons.) and the Dr.Tech. degree (with Commendation) in automation engineering from the Tampere University of Technology, Tampere, Finland. He carried out research at the Departamento de Ingenieria Elctrica y Energtica, Santander, the Institute of Hydraulics and Automation, Tampere, and the Mechatronics Research Laboratory, Massachusetts Institute of Technology, Cambridge. In 1999, he joined the Tampere University of Technology and became a Full Professor of factory automation in 2006. He has published over 250 original articles in international journals and conference proceedings. His research interest includes the applications of information and communication technologies in the field of factory automation and robotics. He is a member of the IEEE Industrial Electronics Society. He is a member of several editorial boards. He was the Deputy Chair of the IEEE/ES Technical Committee on Industrial Cyberphysical Systems from 2019 to 2022. He served as an Associate Editor for IEEE TRANSACTIONS ON INDUSTRIAL INFORMATICS in 2006 and from 2012 to 2022 and a Technical Editor for IEEE/ASME TRANSACTIONS ON MECHATRONICS from 2015 to 2016.